\documentclass{article}
\usepackage{spconf,amsmath,graphicx,boldline,subfig,tabularx,kotex,amsmath}
 
\usepackage{xcolor}

\title{Frame-level speaker embeddings for text-independent speaker recognition and analysis of end-to-end model}
\name{Suwon Shon, Hao Tang, James Glass}
\address{Computer Science and Artificial Intelligence Laboratory \\
Massachusetts Institute of Technology \\
Cambridge, MA 02139 USA \\
{\small\texttt{\{swshon,haotang,glass\}@mit.edu}}}
\begin{document}
%
\maketitle
\begin{abstract}
In this paper, we propose a Convolutional Neural Network (CNN) based speaker recognition model for extracting robust speaker embeddings. The embedding can be extracted efficiently with linear activation in the embedding layer. To understand how the speaker recognition model operates with text-independent input, we modify the structure to extract frame-level speaker embeddings from each hidden layer. We feed utterances from the TIMIT dataset to the trained network and use several proxy tasks to study the networks ability to represent speech input and differentiate voice identity.
We found that the networks are better at discriminating broad phonetic classes than individual phonemes. In particular, frame-level embeddings that belong to the same phonetic classes are similar (based on cosine distance) for the same speaker. The frame level representation also allows us to analyze the networks at the frame level, and has the potential for other analyses to improve speaker
recognition.
\end{abstract}
\begin{keywords}
speaker recognition, embedding, frame-level representation,text-independent \end{keywords}
\section{Introduction}
\label{sec:intro}

Deep neural networks (DNNs) have been actively used in speaker recognition to discriminate speakers' identity.
In most settings, DNNs are used as a replacement for Gaussian mixture models (GMMs) to improve the conventional i-vector approach~\cite{Dehak2011} by having a more phonetically aware Universal Background Model (UBM)~\cite{Richardson2015,DavidSnyder2016,Lei2014}.
Other subsequent method based on DNN were introduced for noise-robust and domain-invariant i-vector~\cite{Ghahabi2014,Pekhovsky2016,Shon2017domain}
However, the process of training the GMM-UBM and extracting i-vectors largely remained the same.

More recently, many studies have begun to explore end-to-end DNN speaker recognition to extract robust speaker embeddings using large datasets as well as data augmentation~\cite{Snyder2018,Nagraniy2017}.
These end-to-end models directly operate on spectrograms, log Mel features, or even waveforms~\cite{Shon2018,Jung2018}. Among the end-to-end approaches, x-vectors have been the most effective for text-independent scenarios~\cite{Snyder2018}. Compared to i-vectors and bottleneck feature-based i-vectors, x-vectors have achieved better results by taking advantage of data augmentation with noise and reverberation. Due to neural networks large learning capacity, data augmentation has been shown to be a cheap and effective approach to improve performance and robustness. The gap between x-vectors and i-vectors is expected to widen as the amount of data increases and end-to-end networks continue to be improved. 

The i-vector approach is based on the assumption that each individual mean vector in a GMM is a shift from a mean vector of the UBM, and that the shifts of all the means are controlled by a single vector, the i-vector.
The model has been studied extensively and is well understood~\cite{Dehak2011}. 
In contrast, it is difficult to understand why and how speaker embedding networks work, which hinders the development of better end-to-end speaker recognition models.

In this paper, we introduce a speaker embedding extracted from a 1-dimensional convolution and linear activation from an end-to-end model. The use of linear activation is inspired by previous studies~\cite{Zhang2014,vesely2011convolutive}, where reducing non-linearities has been shown to improve performance.
The embeddings are compared to two strong baselines, x-vectors and an approach based on the VGG network. We then analyze the networks behavior by modifying the network structure and extracting frame-level representations from the hidden layers. We feed utterances from the TIMIT dataset into the model and monitor the behavior of the representations at different training epochs.
We hypothesize that the networks' ability to recognize speakers
is based on how the phonemes are pronounced and that the networks pay more attention to certain phonemes or broad class than others.
For text-independent input, since it is unlikely that the same set of phonemes appeared in both the enrollment and test utterances, we believe the speakers' identity is less likely to be decided at the phonetic level but more likely at a higher level based on the phonetic classes. Identifying speakers at the broad-class level allows the networks to assess the speaker's voice even without  the presence of the exact same phoneme.

To verify this hypotheses, we conduct phoneme recognition and broad-class classification tasks using frame-level representation of speaker embeddings, and then we visualize and analyze the frame-level cosine similarity measurements from the same and different speaker pairs. From these proxy tasks, we examine how phonetic information is encoded in the network. We also investigate which phoneme or broad-classes are more important for text-independent speaker recognition using frame-level speaker embeddings extracted from TIMIT data. 

\section{speaker embeddings with linear activation}
\label{sec:embeddings}

Previous work has shown that the layer immediately following the statistics pooling layer
performs well in combination with latent discriminant analysis (LDA) and probabilistic LDA (PLDA) as backends~\cite{Snyder2018}.
It is perhaps not surprising that the layer closest to the output layer contains the most discriminative information about the output labels.  We follow a similar approach to analyze our end-to-end speaker recognition model.

\begin{figure}[t]
  \centering
  {\includegraphics[width=0.3\textwidth]{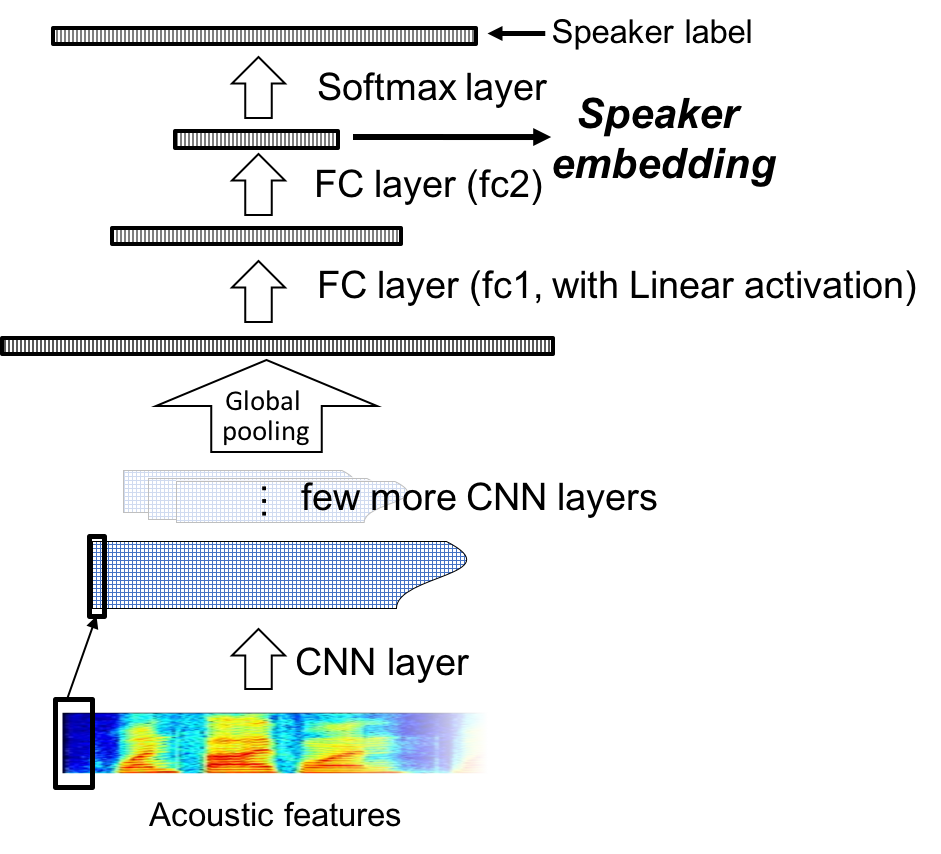}}
  \caption{DNN structure to extract speaker embeddings}
  \label{fig:structure}
\end{figure}

Our network structure is similar to the VGG network but we use 1-dimensional convolutions to consider all frequency bands at once. The network structure consists of 4 1d-CNN (i.e., filters of size 40$\times$5, 1000$\times$7, 1000$\times$1, 1000$\times$1 with strides 1, 2, 1, and 1 and numbers of filters 1000, 1000, 1000, and 1500) with two fully connected (FC) layers (of size 1500 and 600) as shown in Figure~\ref{fig:structure}. We use statistics pooling as in~\cite{Snyder2018}. We use a 1-d CNN instead of a 2-d CNN commonly used for computer vision, because, unlike images, a spectrogram carries different meanings in each axis, namely time versus frequency.  A person's voice can be shifted in time but it cannot be shifted in frequency. For this reason, the width of the 1-d CNN is the entire frequency axis.

We use 40-dimensional Mel-Frequency Cepstral Coefficients (MFCCs) with the standard 25ms window size and 10ms shift to represent the speech signal. The features are normalized to have zero mean. We use the Voxceleb1 development dataset, including 1,211 speakers and 147,935 utterances, to train the networks. The Voxceleb1 test set has 18,860 verification pairs for each positive and negative test, i.e., a combination of 4,715 utterances from 40 speakers not included in the development set. The networks are trained from random initialization. The SGD learning rate is 0.001, and is decayed by a factor of 0.98 after every 50,000 updates.

\begin{table}[t]
\centering
\caption{
Results on the Voxceleb1 test set.
ReLU is applied after every layer.}
\label{tab:withRelu}
\resizebox{0.45\textwidth}{!}{%
\begin{tabular}{l|c|c|c}
\hline
 & \phantom{..}EER\phantom{..} & \phantom{.}DCF$_{p=0.01}$ \phantom{.}& DCF$_{p=0.001}$ \\ \hline
fc1 (LDA+PLDA) & \bf{6.2} & \bf{0.53} & 0.70 \\ \hline
fc2 (LDA+PLDA) & 6.9 & 0.55 & \bf{0.65} \\ \hline
\end{tabular}%
}
\end{table}

\begin{table}[t]
\centering
\caption{
Results on the Voxceleb1 test set.
ReLU is applied after every layer except after fc1.}
\label{tab:withoutRelu}
\resizebox{0.45\textwidth}{!}{%
\begin{tabular}{l|c|c|c}
\hline
 & \phantom{..}EER\phantom{..} & DCF$_{p=0.01}$\phantom{.} & DCF$_{p=0.001}$ \\ \hline
fc1 (LDA+PLDA) & 6.2 & 0.51 & 0.69 \\ \hline
fc2 (LDA+PLDA) & \bf{5.9} & \bf{0.50} & \bf{0.62} \\ \hline
\end{tabular}%
}
\end{table}

Our first experiments use ReLU nonlinearities after each layer in the network. Results are shown in Table~\ref{tab:withRelu}. We find similar results as in~\cite{DavidSnyder2016} that the vectors from the first fully connected layer (fc1) have better performance than those from the second fully connected layer (fc2). 
 
We remove the nonlinear activation function before the last hidden layer (fc2), and the results are shown in Table~\ref{tab:withoutRelu}. From the result, we observe better performance when removing the activation function, and this observation has also been made in previous studies~\cite{vesely2011convolutive, Zhang2014}. We subsequently use the vectors from the fc2 layer as speaker embeddings.

\begin{table*}[ht]
\centering
\caption{
Three recent speaker embedding approaches.
}
\label{tab:comparison_approach}
\resizebox{0.75\textwidth}{!}{%
\begin{tabular}{|c|c|c|c|}
\hline
 & x-vector \cite{DavidSnyder2016} & VGG \cite{Nagraniy2017} & Ours \\ \hline
Input for training & MFCC & \begin{tabular}[c]{@{}c@{}}Spectrogram \\ with fixed length(3sec)\end{tabular} & \begin{tabular}[c]{@{}c@{}}MFCC \\ with fixed length (2sec)\end{tabular} \\ \hline
Input normalization & CMN & CMVN & CMN \\ \hline
Structure & TDNN & 2d-CNN (VGG-M) & 1d-CNN \\ \hline
Parameters & 4.4m & 64m & 13m \\ \hline
Global Pooling & \begin{tabular}[c]{@{}c@{}}Statistics \\ \end{tabular} & Average & Statistics \\ \hline
Embedding layer & First fully connected layer & Last fully connected layer & Last fully connected layer \\ \hline
Nonlinearity & All layers & All layers & \begin{tabular}[c]{@{}c@{}}All layers except\\before embedding layer\end{tabular} \\ \hline
Embedding Dimension & 512 & 1024 & \begin{tabular}[c]{@{}c@{}}600\\\end{tabular} \\ \hline
\begin{tabular}[c]{@{}c@{}}Backend\\ Processing\end{tabular} & \begin{tabular}[c]{@{}c@{}}Zero-mean norm.+LDA\\ +length norm.+PLDA\end{tabular} & \begin{tabular}[c]{@{}c@{}}Euclidean Distance\\ with Siamese network\end{tabular} & \begin{tabular}[c]{@{}c@{}}Zero-mean norm.+LDA\\ +length norm.+PLDA\end{tabular} \\ \hline
\end{tabular}%
}
\end{table*}

The differences between recent speaker embedding approaches are summarized in Table~\ref{tab:comparison_approach}. Using the same setting as in Table~\ref{tab:withoutRelu}, we compare the speaker embeddings with i-vectors, x-vectors, and the approach based on the VGG network. Results are shown in Table~\ref{tab:eer_compare}. We augment the dataset as in \cite{Snyder2018} with reverberation and different noise types, such as babble noise and background music. The number of utterances is 147,935 and we augment an additional 140,000 utterances. Without data augmentation, the proposed speaker embedding method is slightly worse than i-vectors but significantly outperforms the VGG approach and x-vectors.  We find that the i-vectors in \cite{Nagraniy2017} are worse because they use a small number of Gaussian components for the GMM-UBM. After using 2,048 components, the i-vectors perform the best. With data augmentation, the EER improves by 15\% for x-vectors but is still worse than i-vectors. Our embeddings also benefit from data augmentation and is able to match the i-vector results.

\begin{table}[t]
\centering
\caption{
Results on the
Voxceleb1 test set.
Systems trained with data augmentation are labeled with *.}
\label{tab:eer_compare}
\resizebox{0.45\textwidth}{!}{%
\begin{tabular}{l||c|c|c}
\hlineB{2}
 &\phantom{..} EER \phantom{..} & \phantom{..}DCF$_{p=0.01}\phantom{..}$ & DCF$_{p=0.001}$ \\ \hlineB{2}
i-vector & \textbf{5.4 }& \textbf{0.45} & 0.63 \\ \hline
i-vector* & 5.5  & 0.48 & \textbf{0.61} \\ \hlineB{2}
VGG~\cite{Nagraniy2017} & 7.8 & 0.71 & - \\ \hline
x-vector (Cosine) & 11.3 & 0.75 & 0.81 \\ \hline
x-vector (PLDA) & 7.1 & 0.57 & 0.75 \\ \hline
x-vector* (Cosine) & 9.9 & 0.69 & 0.85 \\ \hline
x-vector* (PLDA) & 6.0 & 0.53 & 0.75 \\ \hline
fc2 (Cosine) & 7.3 & 0.56 & 0.64 \\ \hline
fc2 (PLDA) & 5.9 & 0.50 & \textbf{0.62} \\ \hline
fc2* (Cosine) & 7.0 & 0.58 & 0.68 \\ \hline
fc2* (PLDA) & \textbf{5.3}& \textbf{0.45} & 0.63 \\ \hlineB{2}
\end{tabular}%
}
\end{table}

\section{Frame-level representation of speaker embedding and its analysis}

Conventional speaker recognition approaches, such as i-vectors, require many steps that are carefully designed for learning a robust representation of speaker identity from acoustic features. Representations learned from deep networks, however, are optimized for this particular task purely from data~\cite{Lecun2015}.
The output vectors produced by the intermediate layers, the hidden representation, could be the key to understand what end-to-end speaker recognition model implicitly learns from voice inputs.
We adopt the approach used in \cite{AKBLG2017, Belinkov2017} where several proxy tasks are used to analyze the hidden representations.
In speaker recognition, several studies have assessed which phoneme contributes the most to discriminating speakers~\cite{Eatock1994}, and which phonetic classes are more important than others~\cite{Antal2006}.
Both papers conclude that vowels and nasals
provide the most useful information for identifying speakers. However, the experiments are limited to a single phoneme or a single phonetic class,
so it is difficult to draw similar conclusions when the networks can make use of an entire utterance. Wang et al.~\cite{Wang2017} analyze speaker embeddings in a text-dependent speaker recognition system, and use an approach similar 
to~\cite{AKBLG2017}.  However, text-dependent speaker recognition is easier to analyze than the text-independent case because the enrollment and test utterances always have the same phoneme statistics.
We aim to understand how the embedding representation encodes phoneme information at the frame level when text-independent input is given.

In this section, we analyze what phonetic information is encoded in the end-to-end speaker recognition model and how they capture and discriminate between talkers given text-independent input.
We use phoneme recognition and phonetic classification as proxy tasks
and monitor the behavior over the course of training to answer these questions.
We also identify phonemes that are critical for text-independent speaker verification. 
For the proxy tasks, we assume that the ability of a classify to predict a certain property depends on how well the property has been encoded in the representation (as in previous studies~\cite{AKBLG2017,Belinkov2017,Wang2017}).  Poor accuracy does not necessarily mean the information is not present however.

\subsection{Frame-level speaker embeddings}

To obtain frame-level representations, we need to modify the structure of our models.
For training, we substitute the statistics pooling layer to be an average pooling layer. After the modification, the EER increased from 7.0\% to 8.4\% using cosine similarity, and from 5.3\% to 6.0\% with PLDA.
After training, we moved the average pooling layer to be after the fc2 layer but just before ReLU activation function. This modification does not change the final results, because multiplying the average with a matrix is the same as the average of the individual vectors multiplied by the matrix.
The output vectors produced by all layers before the average pooling yields frame-level representations not only for CNN layers but also for FC layers.
Specifically, suppose $u$ is a segment-level representation (i.e. a speaker embedding extracted at the fc2 layer in Figure~\ref{fig:structure}).
The embedding $u$ can be represented as the frame-level representation $\vec{u}_t$ which is extracted from the modified model structure as
$u = (1/T) \sum_{t}^{T} \vec{u}_t$ where $T$ is the number of frames in the utterance.
We extracted this frame-level representation
at all CNN and FC layers for the proxy tasks.

\begin{figure}[t]
  \centering
  {\includegraphics[width=0.5\textwidth]{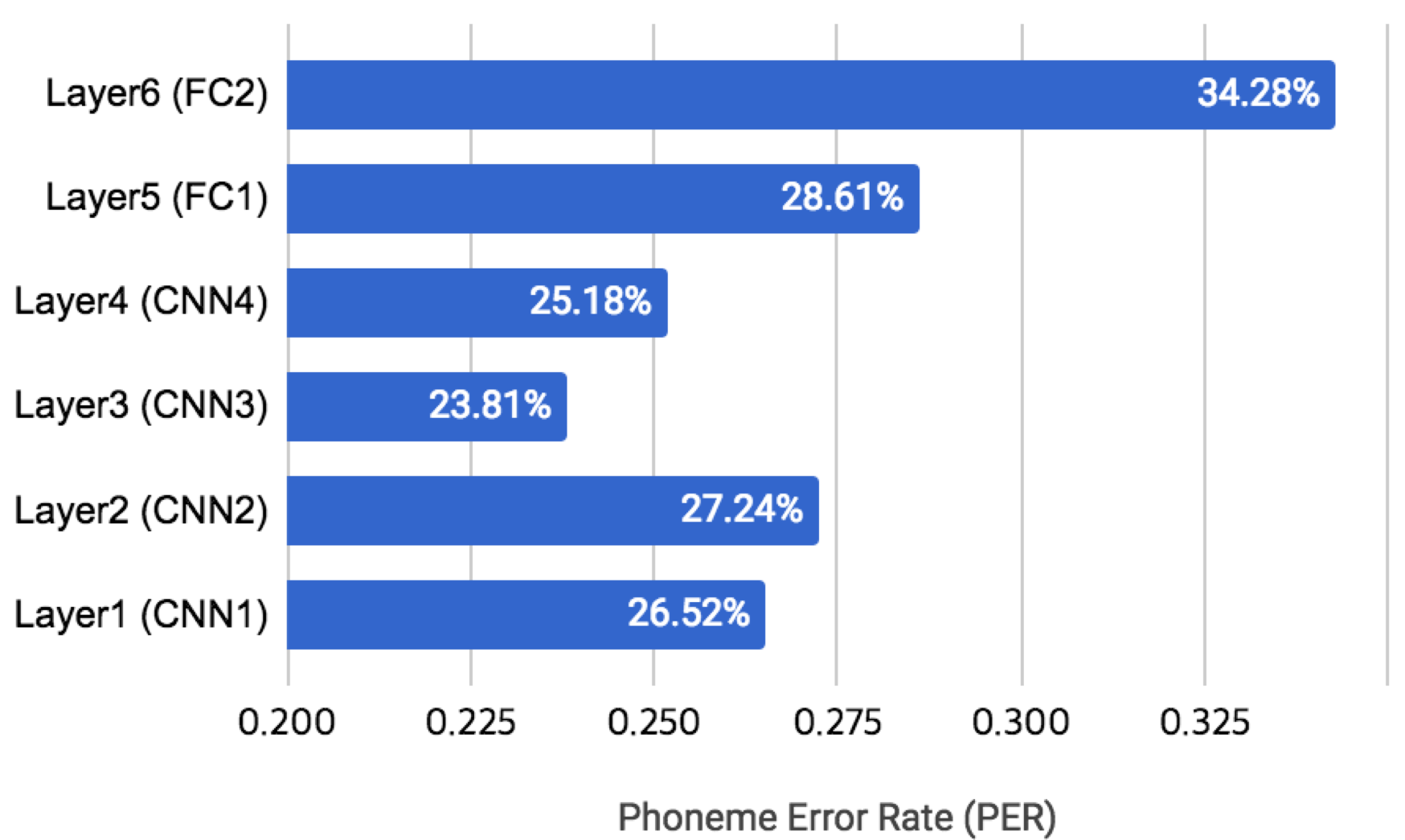}}
  \caption{Phoneme error rates of segmental models on the TIMIT test set using frame-level representations from different layers. Lower is better.}
  \label{fig:per_layer}
\end{figure}

\begin{figure}[t]
  \centering
  {\includegraphics[width=0.5\textwidth]{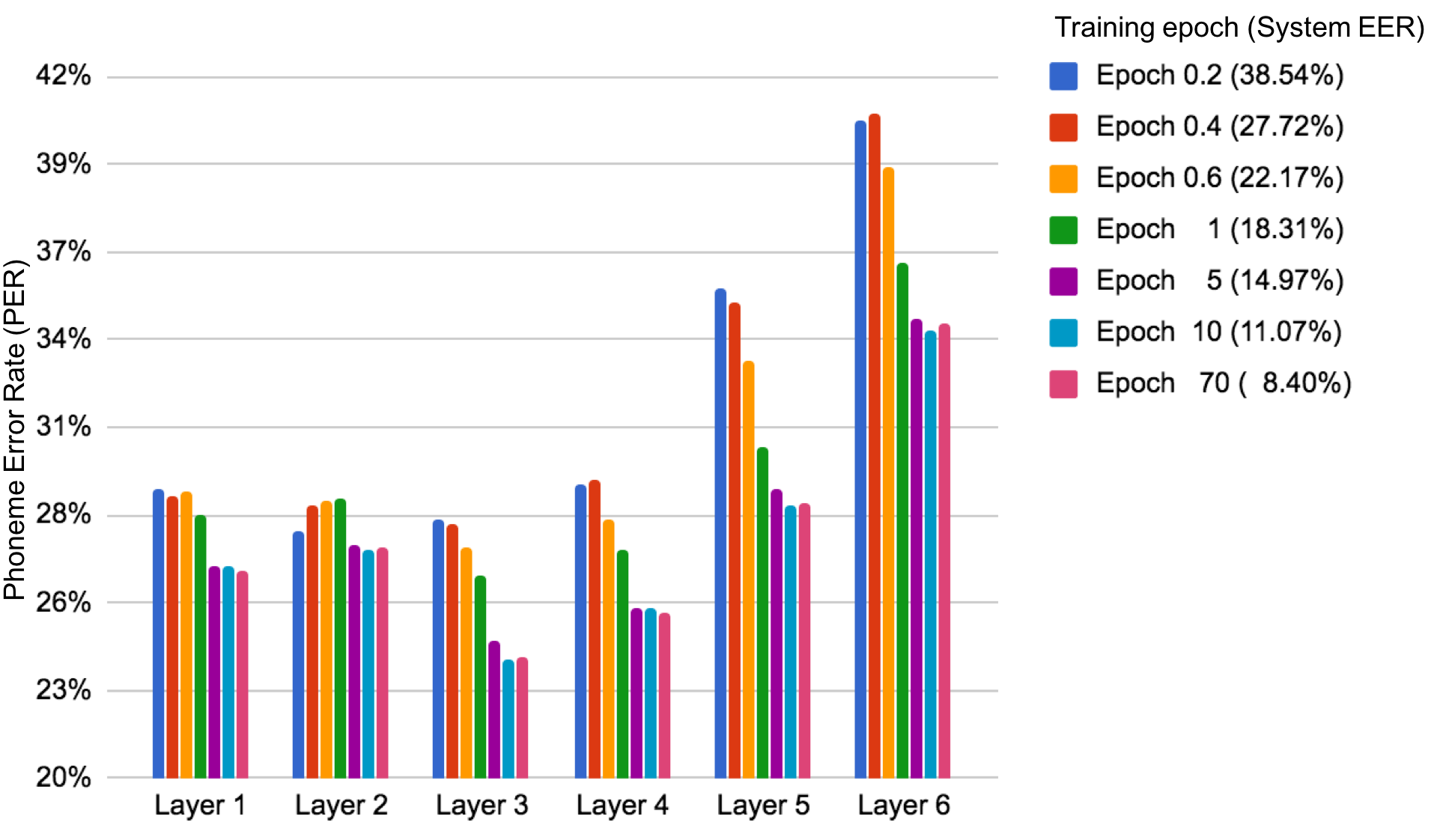}}
  \caption{Phoneme error rates of segmental models on the TIMIT test set using frame-level representation from different layers over the course of training. Lower is better.}
  \label{fig:per_epoch}
\end{figure}

\begin{table}[ht]
\centering
\caption{TIMIT broad phonetic classes.}
\resizebox{0.3\textwidth}{!}{%
\begin{tabular}{|l|l|}
\hline
Class & Symbol \\ \hline
Affricate & jh, ch \\ \hline
Closures & bcl,dcl,gcl,pcl,tck,kcl \\ \hline
Fricative & s,sh,z,zh,f,th,v,dh \\ \hline
Nasals & m,n,ng,em,en,eng,nx \\ \hline
\begin{tabular}[c]{@{}l@{}}Semivowels \\ and Glides\end{tabular} & l, r, w, y, hh, hv, el \\ \hline
Vowels & \begin{tabular}[c]{@{}l@{}}iy, ih, eh, ey, ae, aa, aw, \\ ay, ah, ao, oy, ow, uh, uw,\\ ux, er, ax, ix, axr, ax-h\end{tabular} \\ \hline
Stops & b, d, g, p, t, k, dx, q \\ \hline
others & \begin{tabular}[c]{@{}l@{}}pause(pau), epenthetic silence(epi), \\start and end silence (h)\end{tabular}
 \\ \hline
\end{tabular}%
}
\label{tab:broad-class}
\end{table}

\begin{figure}[ht]
  \centering
  {\includegraphics[width=0.5\textwidth]{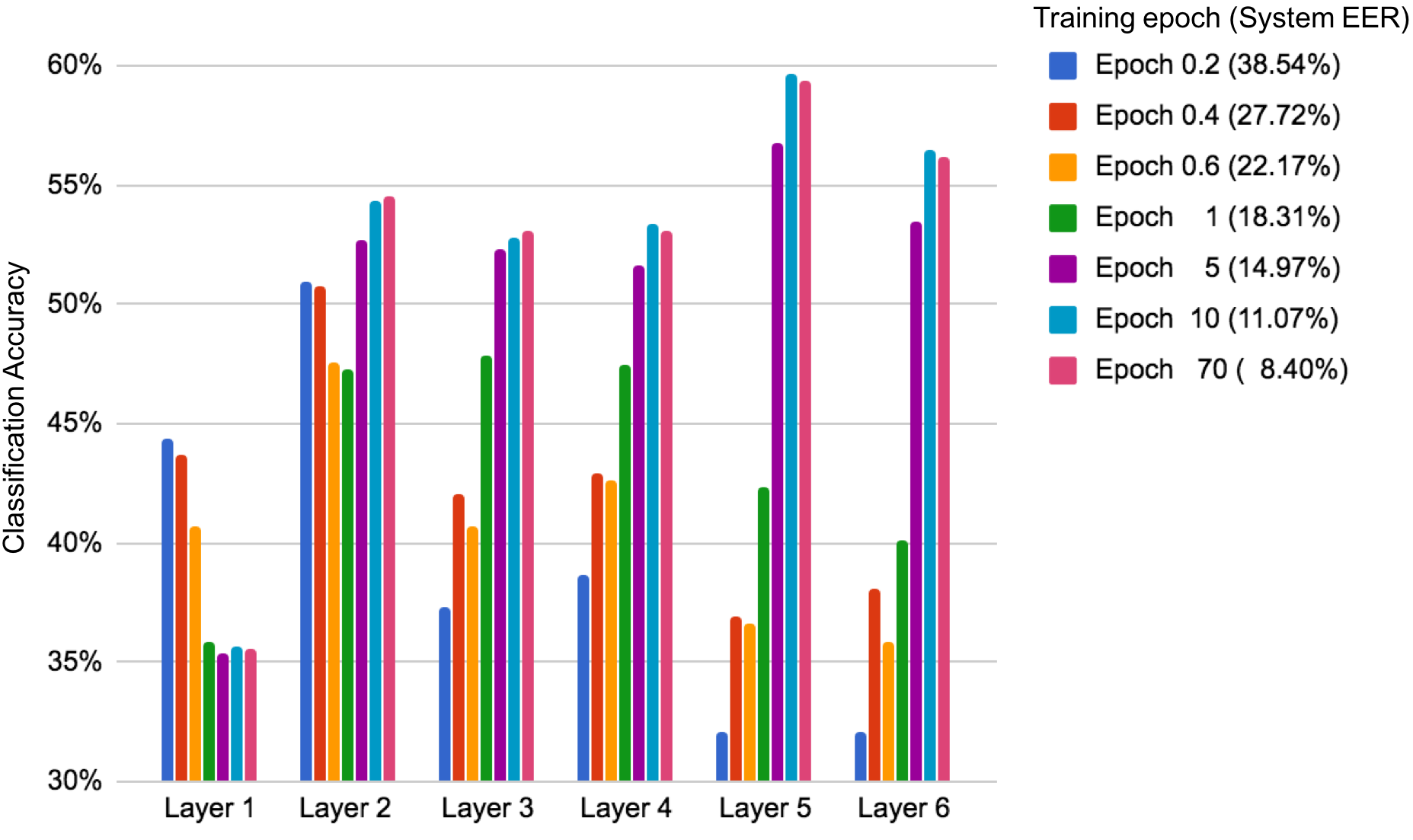}}
  \caption{Phonetic class classification accuracy on TIMIT Test set using features from each layer over the course of training. Higher is better.}
  \label{fig:broadclass_classification}
\end{figure}

\subsection{Phoneme Recognition}

Given the trained model with the modified structure, we use TIMIT utterances to examine the ability of the speaker recognition system at distinguishing phonemes.
The first CNN layer produces an output every 10ms while subsequent layers produce an output every 20ms because of the second layer stride by 2 that reduces the effective analysis rate to 20ms.
The stride should not affect the analyses, because previous work has shown that similar, if not better, results can be achieved with one-fourth of the original frame rate \cite{T+2017}.

We train a phoneme recognizer with the frame-level representation extracted from each of the layers and analyze their performance.
We use an end-to-end discriminative segmental model \cite{T+2017} with a 2-layer LSTM model as our phoneme recognizer.
The input to the LSTM consists of frame-level representations from different layers from the end-to-end speaker recognition model.
In each layer, the output vectors of the LSTM are sub-sampled by half.
The final segment scores are based on the output vectors of the LSTM within the segment and the duration of the segments (the FCB feature set in \cite{T2017}). We allow a segment to have a maximum duration of 120 frames.
The segmental models are trained with marginal log loss \cite{TWGL2016} for 20 epochs with vanilla stochastic gradient descent (SGD), a step size of 0.1. The batch size is one utterance, and the gradients are clipped to norm 5.

Figure~\ref{fig:per_layer} shows the phoneme error rate (PER) for different layers, and indicates that the embeddings from higher layers give higher PERs. Judging from the PERs over the course of training, the training error, in general, stops improving after epoch five.
At layer 6 for example, the PER plateaus at 34\%. From this observation, the frame-level representation contains less information about phonemes, and the phoneme identify
does not seem to be important for discriminating speakers.

\begin{figure}[ht]
  \centering
  \subfloat[Layer 1]{\includegraphics[width=0.257\textwidth]{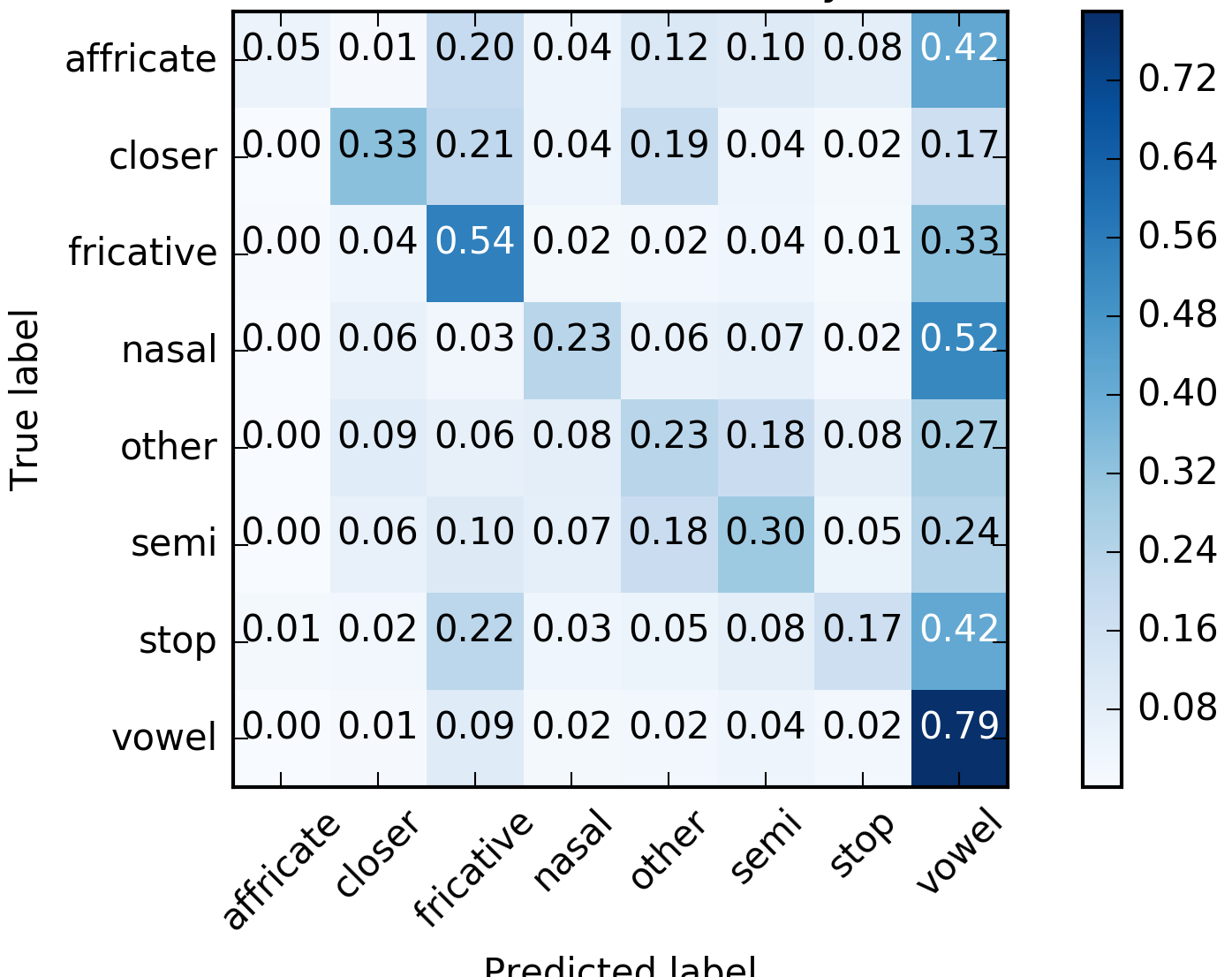}}
  \subfloat[Layer 6]{\includegraphics[width=0.25\textwidth]{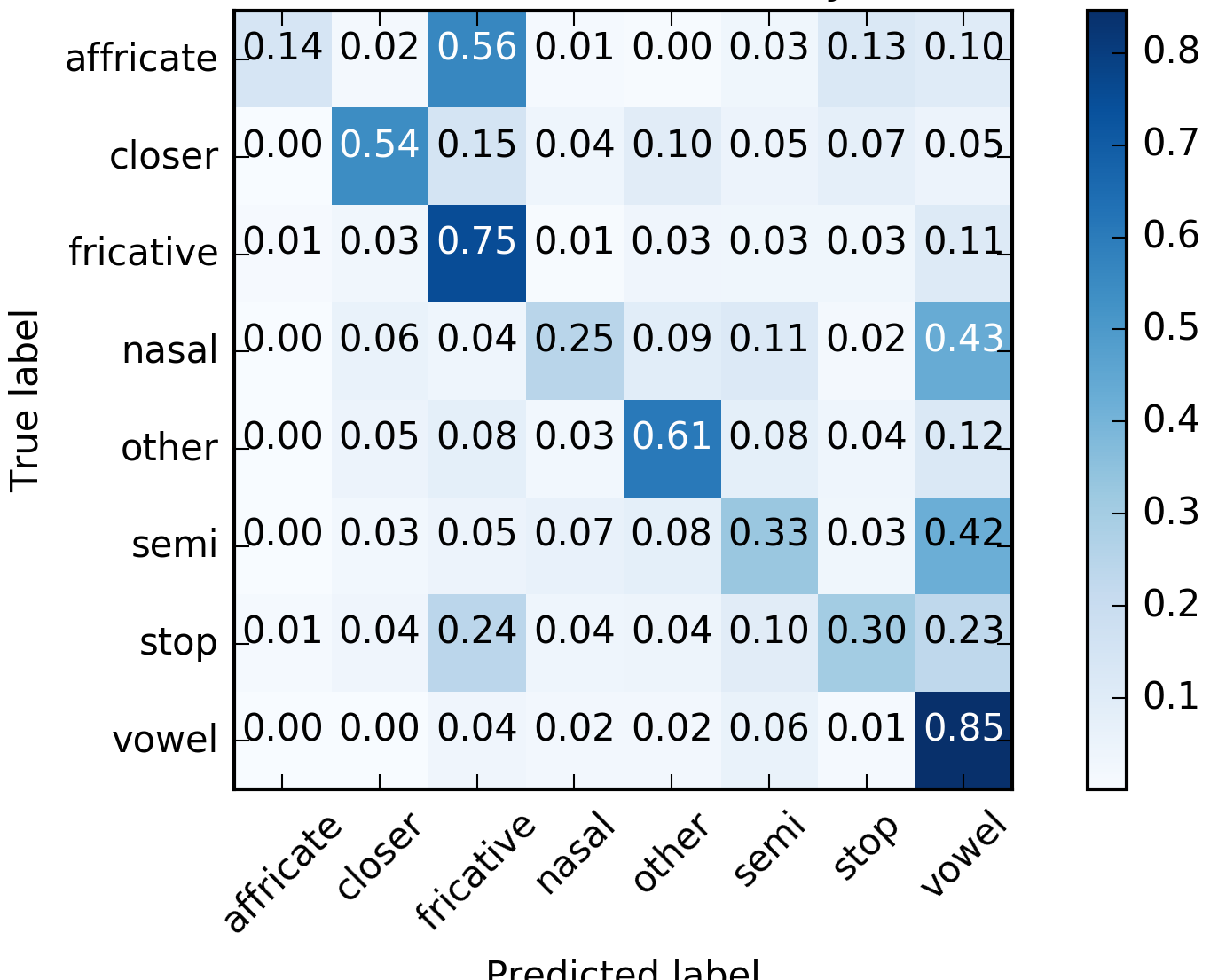}}
  \caption{
  Confusion matrix for broad-class phonetic classification
  comparing the features produced by layer 1 and
  layer 6 at epoch 70.
  }

\label{fig:broadclass_confusion}
\end{figure}

\begin{figure}[ht]
  \centering
  \subfloat[Layer 1]{\includegraphics[width=0.2\textwidth]{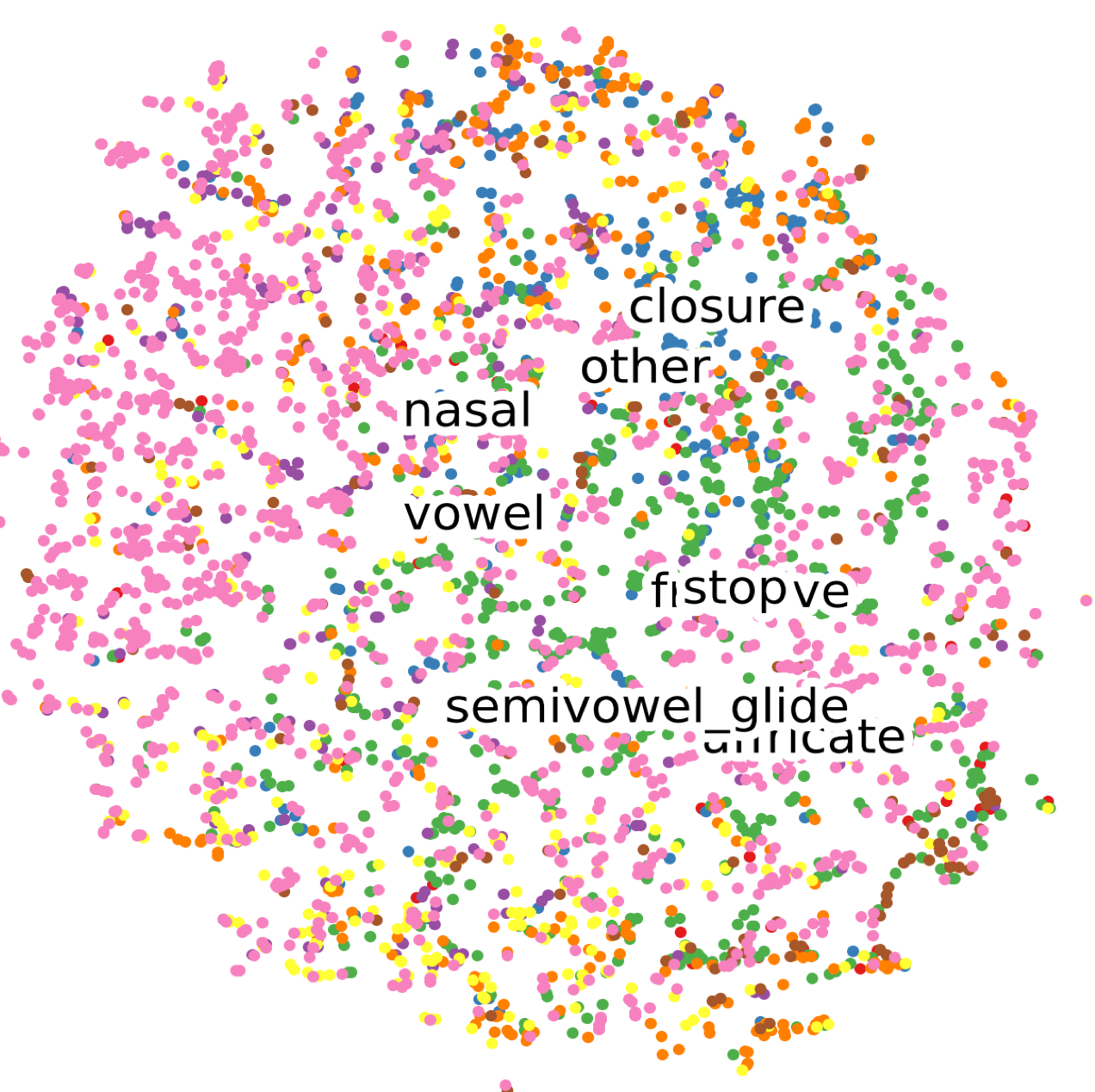}}
  \hfill
  \subfloat[Layer 6]{\includegraphics[width=0.25\textwidth]{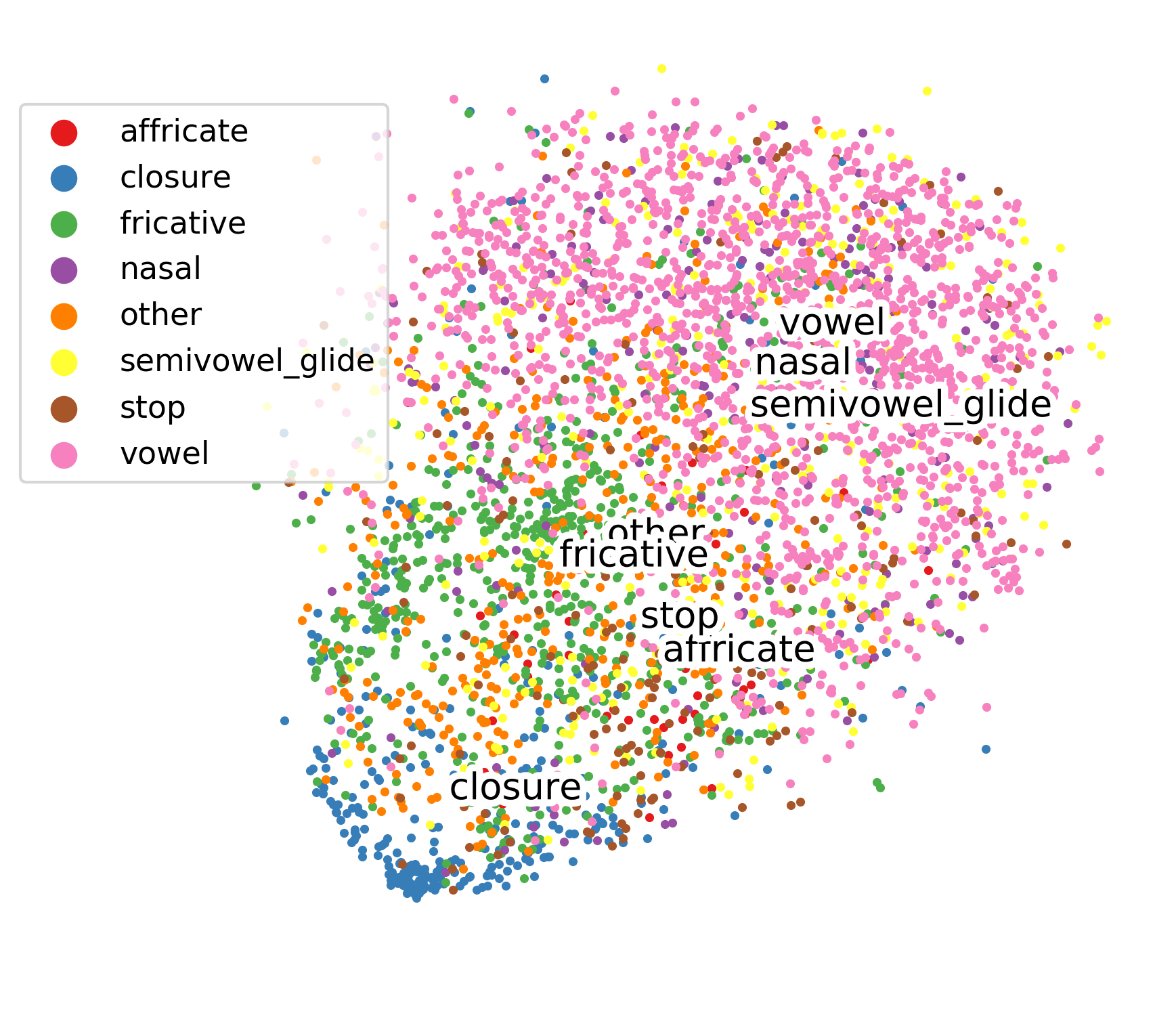}}
  \caption{Low-dimensional t-SNE projection of speaker embeddings from layer 1 and 6 at epoch 70 using utterances from the TIMIT core test set. The broad-class label is printed in the figure.}
  \label{fig:broadclass_cluster}
\end{figure}

\subsection{Broad-Class Phonetic Classification}
For broad-class phonetic classification, we collect all phoneme segments in the TIMIT dataset based on the ground truth segmentation.
The broad phonetic classes we use are shown in Table~\ref{tab:broad-class}.
The segment embedding of each segment is computed
by averaging the frame-level embedding
obtained from the trained speaker recognition system.
We create a naive classifier by averaging the
embeddings of the same phonetic classes.
Specifically, we compute a vector $U^b = \sum_{i \in P_b} u_i / |P_b|$ for phonetic class $b$, where $P_b$ contains the segments of class $b$, and $u_i$ is the segment embedding of segment $i$ by averaging the frame embedding in segment $i$.
Given a new segment $j$, we compute its segment embedding $u_j$ by averaging the frame embeddings computed from the trained network.
Classification can be done with $\operatorname*{argmax}_b (\cos(U^b, u_j))$.
As shown in Figure~\ref{fig:broadclass_classification}, the system at the early stage of training does not distinguish the phonetic classes well, and is worst in the higher layer.
After training, the model learns to distinguish phonetic classes well.
In particular, the representation in the higher layers performs significantly better than the ones in the lower layers.

The confusion matrix of the broad phonetic classes is shown in Figure~\ref{fig:broadclass_confusion}.
The diagonal of the confusion matrix shows that the accuracy of the embedding from layer 6 performs better the one from layer 1.
An interesting observation is that some categories are still confusable at layer 6. For example, affricates and stops are predicted as fricatives, and nasals and semivowels are predicted as vowels.
This suggests that the model might be classifying segments into even broader categories, such as obstruents and sonorants.
This phenomenon is also observed in the t-SNE plots of the phoneme segments shown in Figure~\ref{fig:broadclass_cluster}.

\begin{figure}[t]
  \centering
  \subfloat[Example of frame-level cosine similarity between the same speaker's segment-level and frame-level speaker embedding]{\includegraphics[width=0.45\textwidth]{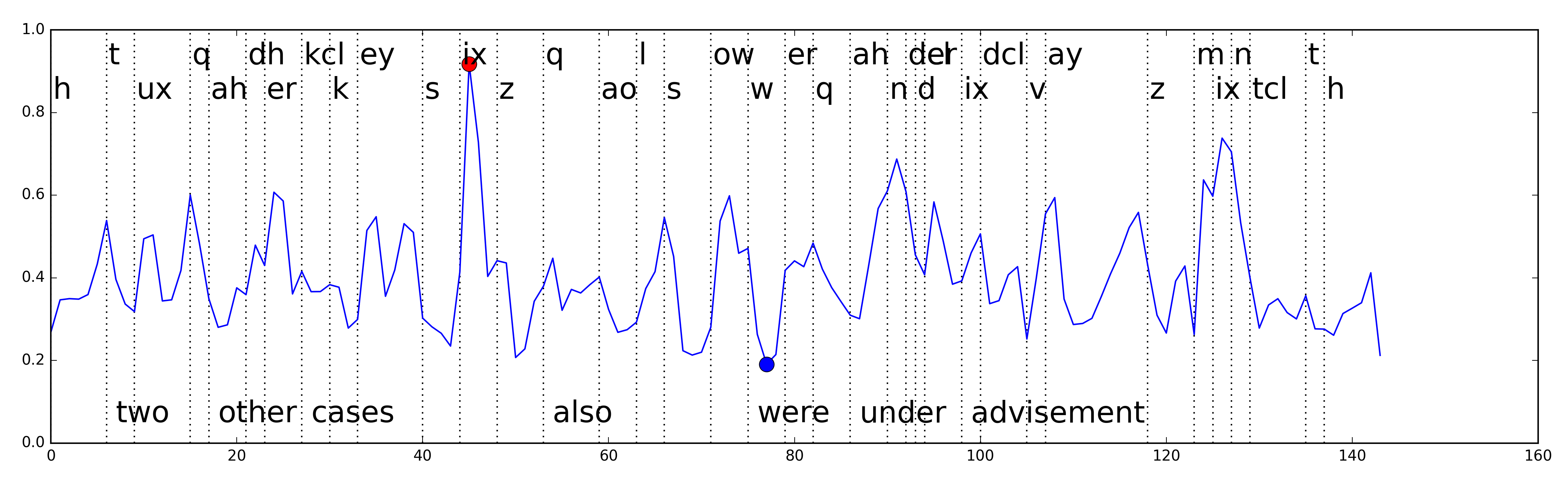}}
  \hfill
  \subfloat[Original phoneme occurrence histogram on TIMIT training set]{\includegraphics[width=0.45\textwidth]{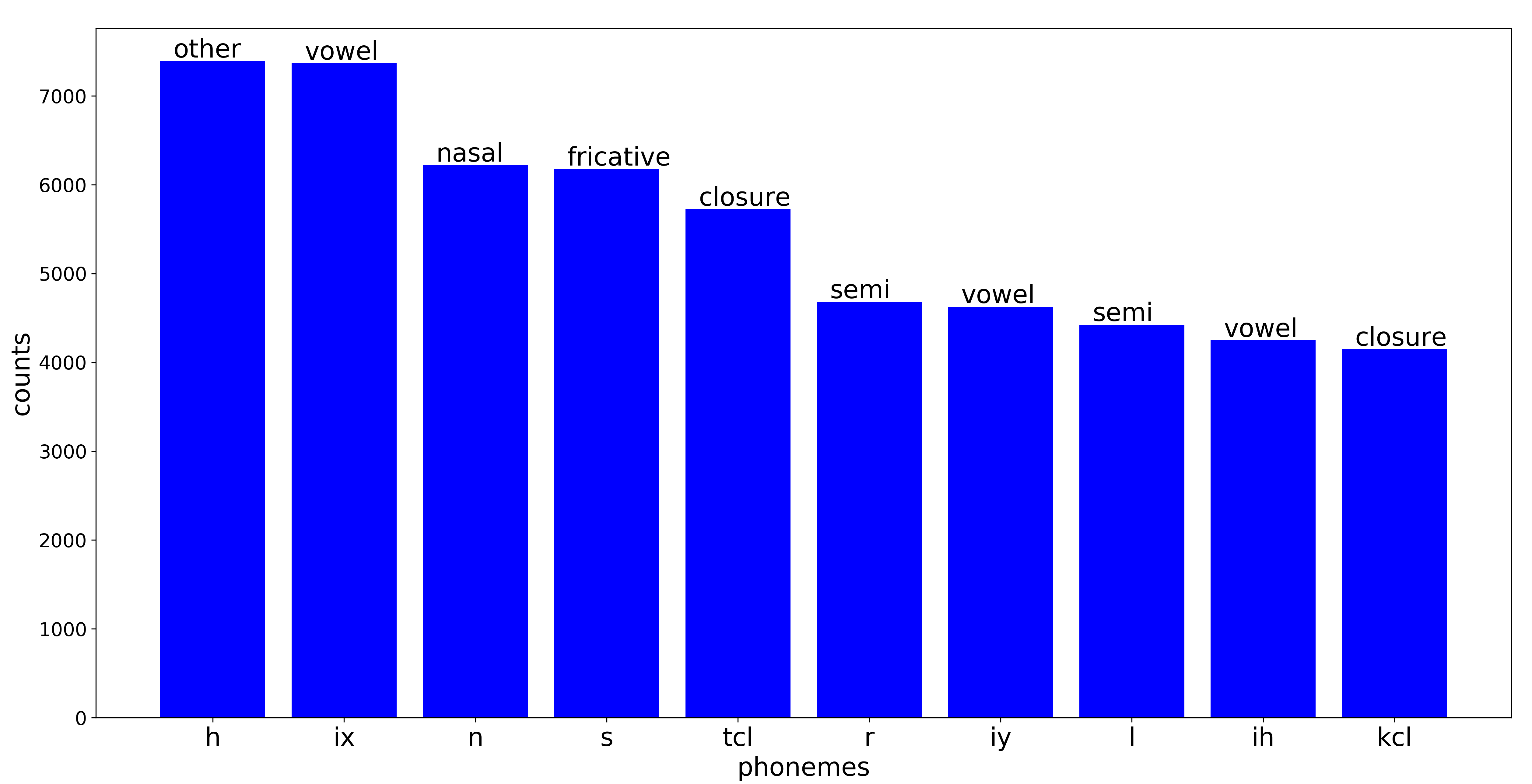}}
  \hfill
  \subfloat[Histogram of highest cosine similarity phonemes in each utterance on TIMIT training set (numbers after words : order from the original histogram )]{\includegraphics[width=0.45\textwidth]{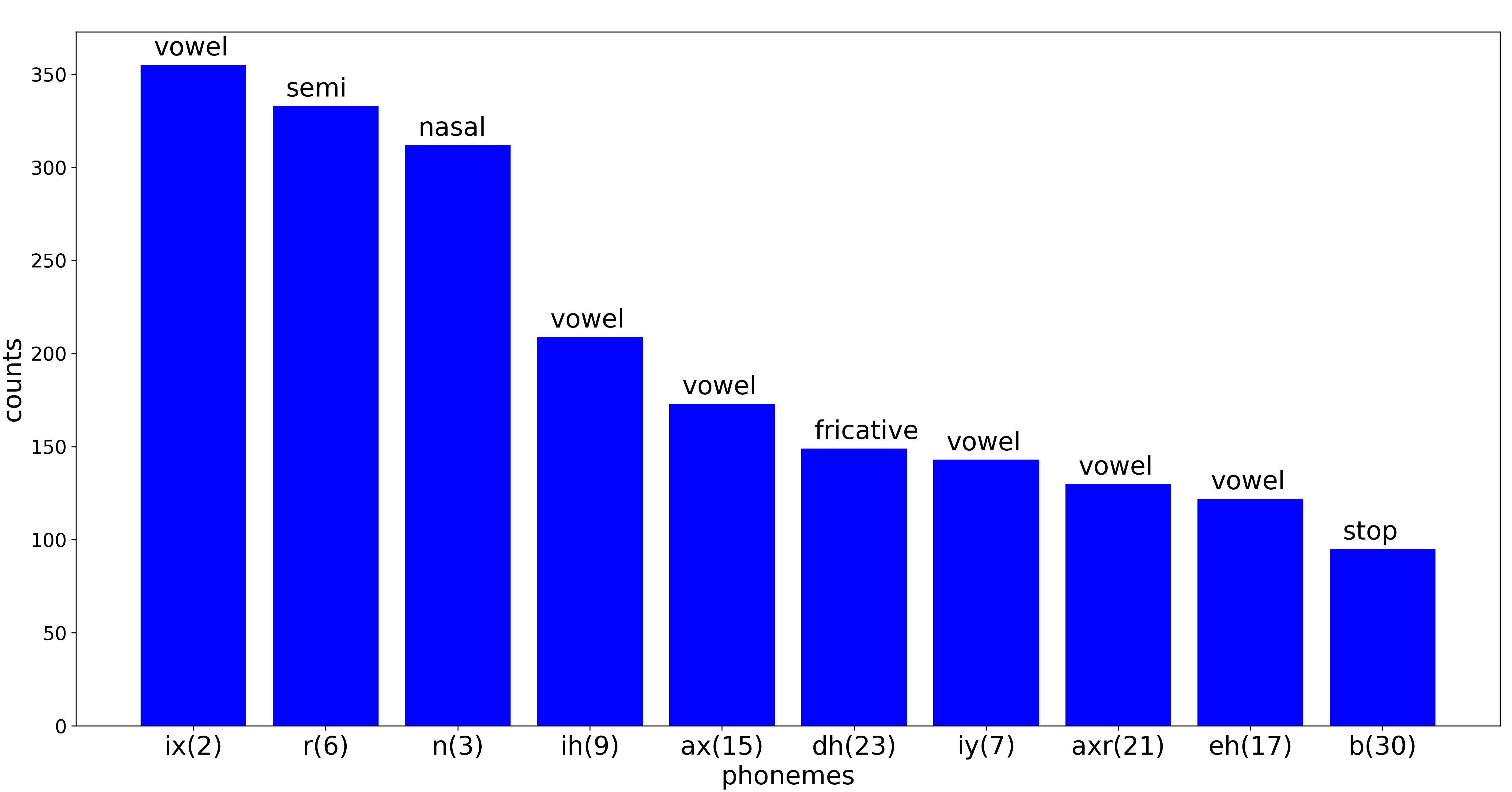}}  \hfill
  \caption{Stats in terms of phoneme}
  \label{fig:critical_phoneme}
\end{figure}

\begin{figure}[b]
  \centering
  \subfloat[Same speaker (speaker id 'fame0' in TIMIT)]{\includegraphics[width=0.4\textwidth]{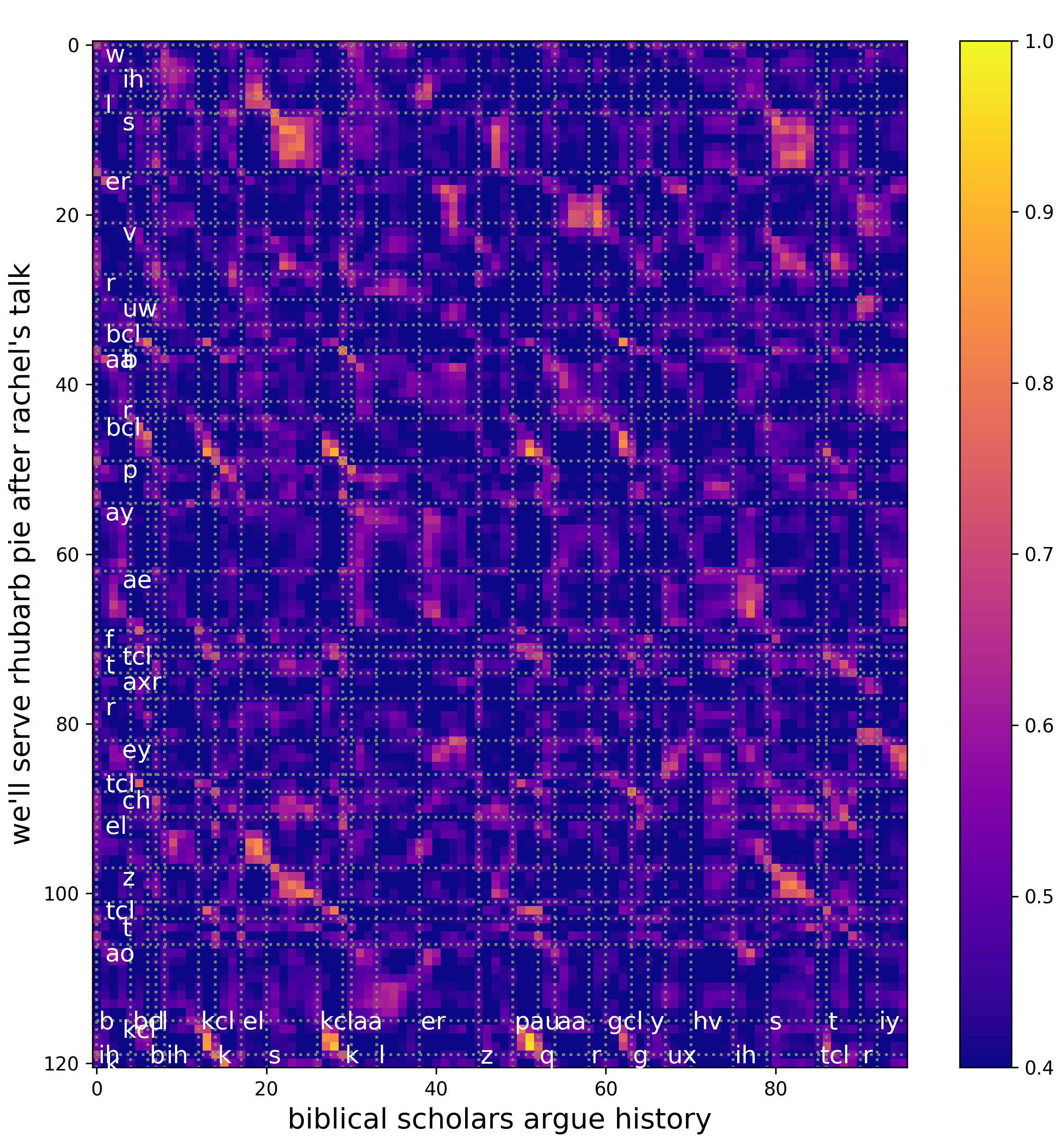}}  \hfill
  \subfloat[Different speakers (speaker id 'faem0' for y-axis and 'mrpc1' for x-axis in TIMIT)]{\includegraphics[width=0.4\textwidth]{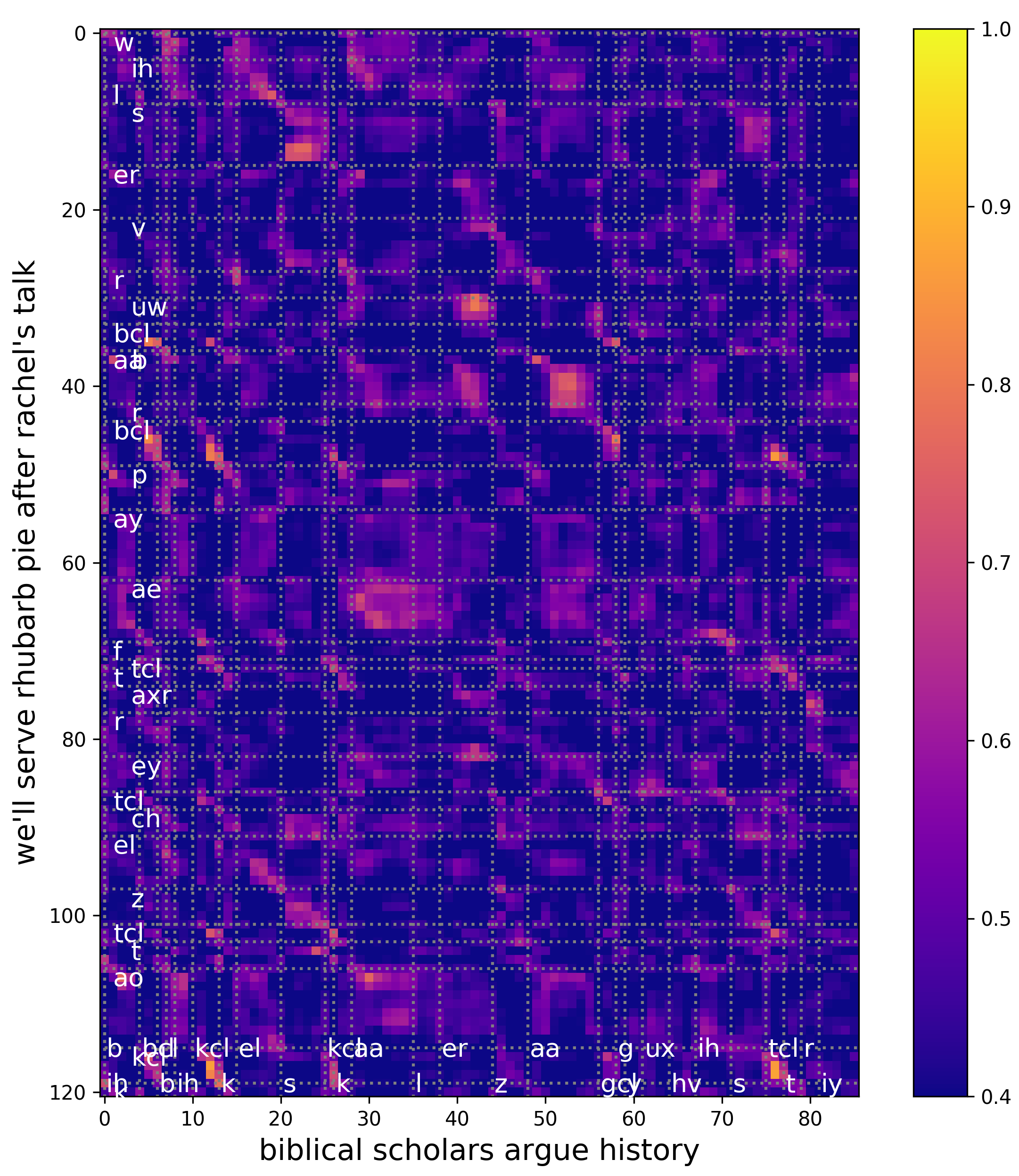}}
  \caption{Frame-level cosine similarity matrix between two sentences spoken by the same speaker
  and by different speakers.}
  \label{fig:similarity_matrix}
\end{figure}

\subsection{Critical phonemes and phonetic classes}
In this section, we analyze cosine similarity at the frame level.
In the TIMIT dataset there are 10 utterances per speaker.
We calculated cosine similarity for all utterances using frame-level speaker embeddings from the modified structure.
For enrollment, for each speaker nine out of ten utterances are used by averaging the segment embeddings to create a single speaker embedding. 
Figure~\ref{fig:critical_phoneme} (a) shows the frame-level cosine similarity with the single speaker embedding. In the example, the phoneme /ix/ shows the highest cosine similarity (blue dot) and
the phoneme /w/ shows the lowest (red dot). Figure~\ref{fig:critical_phoneme} (b)
shows the histogram of
phoneme frequency in the training set,
and (c) shows how often each phonetic class
achieves the highest cosine similarity.
From the histogram, we made similar observation as previous studies~\cite{Eatock1994, Antal2006} that
vowels and nasals are important for discriminating speakers. 
These observations suggests that using an attention mechanism at the pooling layer
may improve the performance of speaker recognition.

Figure~\ref{fig:similarity_matrix} shows the frame-level similarity matrix between two utterances from the TIMIT training set. In (a), the two utterances are spoken by the same speaker and (b) by different speakers. For comparison, the same text content is chosen in both figures. From the figures, the same phoneme shows relatively high similarity even for different speakers. However, in the case of (a) which is spoken by the same speaker, the phonemes that share the same phonetic classes show high similarity. For example, /s/ and /er/ show high similarity with /z/ and /aa/, respectively. From this result, we observe that the network could also potentially measure a speaker's similarity across different phonemes. Interestingly, obstruents, such as fricatives and closures, have relatively high similarity scores even for different speakers.

\subsection{Discussion}
Using both frame-level and segment-level representations, we observe that the representations from higher network layers corresponds to broader classes, while the representation from lower layers correspond to classes that are more specific. Particularly, the representation appears to converge towards obstruent and sonorant categories at higher layers. This behavior suggests that the model computes similarity at a more abstract level than that at the phonetic level. It potentially could provide an advantage when the input is text-independent and short because the model has fewer phonemes to compare with the enrollment data. This is consistent with the results found in the similarity matrix
in Figure~\ref{fig:similarity_matrix}.

\section{Conclusion}
\label{sec:conclusion}

In this paper, we proposed a robust speaker embedding for speaker verification. Embeddings are extracted without non-linear activation and are compared to other approaches to verify their effectiveness. On the Voxceleb1 dataset with only 1.2k speakers, the proposed approach shows superior performance compared to x-vectors. In this framework, there is still room for improvement, such as exploring a larger speaker dataset, a different loss function, such as the angular Softmax loss, or adding an attention layer. We leave these as future work.

From the analysis, we attempt to better understand how the speaker recognition model extracts discriminative embeddings. The analysis provides some insight on the model behavior, and the frame-level analysis provides an important tool to assess the quality of the trained models.

\clearpage

\bibliographystyle{IEEEbib}
\bibliography{strings,refs}

\end{document}